\documentclass[apjl]{emulateapj}
\usepackage{comment}
\usepackage{ifthen}

\newcommand{\forloop}[5][1]%
{%
\setcounter{#2}{#3}%
\ifthenelse{#4}%
	{%
	#5%
	\addtocounter{#2}{#1}%
	\forloop[#1]{#2}{\value{#2}}{#4}{#5}%
	}%
	{%
	}%
}%

\newcommand{\ctbd}[1]{}

\newcommand{\lc}{light curve}
\newcommand{\lcs}{light curves}
\newcommand{\Lc}{Light curve}

\newcommand{\band}[1]{\ensuremath{#1}~band}

\newcommand{\kms}{\ensuremath{\rm km\,s^{-1}}}
\newcommand{\ms}{\ensuremath{\rm m\,s^{-1}}}

\newcommand{\gcmc}{\ensuremath{\rm g\,cm^{-3}}}
\newcommand{\ergscmsq}{\ensuremath{\rm erg\,s^{-1}\,cm^{-2}}}

\newcommand{\teff}{\ensuremath{T_{\rm eff}}}
\newcommand{\logg}{\ensuremath{\log{g}}}
\newcommand{\vsini}{\ensuremath{v \sin{i}}}
\newcommand{\feh}{\ensuremath{\rm [Fe/H]}}

\newcommand{\rsun}{\ensuremath{R_\sun}}
\newcommand{\msun}{\ensuremath{M_\sun}}
\newcommand{\lsun}{\ensuremath{L_\sun}}

\newcommand{\rstar}{\ensuremath{R_\star}}
\newcommand{\mstar}{\ensuremath{M_\star}}
\newcommand{\lstar}{\ensuremath{L_\star}}

\newcommand{\teffstar}{\ensuremath{T_{\rm eff\star}}}
\newcommand{\rhostar}{\ensuremath{\rho_\star}}
\newcommand{\loggstar}{\ensuremath{\log{g_{\star}}}}

\newcommand{\rpl}{\ensuremath{R_{p}}}
\newcommand{\mpl}{\ensuremath{M_{p}}}

\newcommand{\rhopl}{\ensuremath{\rho_{p}}}

\newcommand{\arstar}{\ensuremath{a/\rstar}}
\newcommand{\zrstar}{\ensuremath{\zeta/\rstar}}

\newcommand{\rjup}{\ensuremath{R_{\rm J}}}
\newcommand{\mjup}{\ensuremath{M_{\rm J}}}

\newcommand{\reffigl}[1]{Figure~\ref{fig:#1}}
\newcommand{\refsecl}[1]{\mbox{Section \ref{sec:#1}}}

\newcommand{\reftabl}[1]{Table~\ref{tab:#1}}

\newcommand{\flwof}{\mbox{FLWO 1.2\,m}}

\newcommand{\hatcurCCra}{\ensuremath{20^{\mathrm h}21^{\mathrm m}46.08{\mathrm s}}}                                  
\newcommand{\hatcurCCdec}{\ensuremath{+26{\arcdeg}41{\arcmin}33.5{\arcsec}}}                                 
\newcommand{\hatcurCCtwomass}{2MASS~20214593+2641335}                  
\newcommand{\hatcurCCgsc}{GSC~2163-00549}                              
\newcommand{\hatcurCChd}{HD~340099}                                    
\newcommand{\hatcurCCtassmv}{\ensuremath{10.326\pm0.010}}              
\newcommand{\hatcurCCtassmvshort}{\ensuremath{10.3}}                   
\newcommand{\hatcurCCtassmB}{\ensuremath{10.675\pm0.010}}              
\newcommand{\hatcurCCtassmI}{\ensuremath{9.597\pm0.061}}               
\newcommand{\hatcurCCtassmg}{\ensuremath{10.504\pm0.010}}              
\newcommand{\hatcurCCtassmr}{\ensuremath{10.264\pm0.010}}              
\newcommand{\hatcurCCtassmi}{\ensuremath{10.206\pm0.010}}              
\newcommand{\hatcurCCtwomassJmag}{\ensuremath{9.550\pm0.020}}          
\newcommand{\hatcurCCtwomassHmag}{\ensuremath{9.399\pm0.021}}          
\newcommand{\hatcurCCtwomassKmag}{\ensuremath{9.346\pm0.017}}          
\newcommand{\hatcurLCdip}{\ensuremath{7.6}}                            
\newcommand{\hatcurLCrprstar}{\ensuremath{0.0792\pm0.0019}}            
\newcommand{\hatcurLCbsq}{\ensuremath{0.116_{-0.068}^{+0.102}}}        
\newcommand{\hatcurLCimp}{\ensuremath{0.340_{-0.141}^{+0.119}}}        
\newcommand{\hatcurLCzeta}{\ensuremath{12.74\pm0.10}}                  
\newcommand{\hatcurLCdur}{\ensuremath{0.1712\pm0.0019}}                
\newcommand{\hatcurLCingdur}{\ensuremath{0.0141\pm0.0017}}             
\newcommand{\hatcurLCP}{\ensuremath{2.691548\pm0.000006}}              
\newcommand{\hatcurLCPshort}{\ensuremath{2.6915}}                      
\newcommand{\hatcurLCT}{\ensuremath{2456399.62406\pm0.00063}}          
\newcommand{\hatcurSMEiteff}{\ensuremath{6668\pm52}}                   
\newcommand{\hatcurSMEizfeh}{\ensuremath{0.00\pm0.08}}                 
\newcommand{\hatcurSMEizfehshort}{\ensuremath{0.00}}                   
\newcommand{\hatcurSMEilogg}{\ensuremath{3.95\pm0.10}}                 
\newcommand{\hatcurSMEivsin}{\ensuremath{16.02\pm0.50}}                
\newcommand{\hatcurSMEivmac}{\ensuremath{NULL}}                        
\newcommand{\hatcurSMEivmic}{\ensuremath{NULL}}                        
\newcommand{\hatcurSMEiiteff}{\ensuremath{6820\pm52}}                  
\newcommand{\hatcurSMEiizfeh}{\ensuremath{0.074\pm0.080}}              
\newcommand{\hatcurSMEiizfehshort}{\ensuremath{0.074}}                 
\newcommand{\hatcurSMEiilogg}{\ensuremath{4.10\pm0.06}}                
\newcommand{\hatcurSMEiivsin}{\ensuremath{16.00\pm0.50}}               
\newcommand{\hatcurSMEiivmac}{\ensuremath{NULL}}                       
\newcommand{\hatcurSMEiivmic}{\ensuremath{NULL}}                       
\newcommand{\hatcurTRESgamma}{\ensuremath{15.155\pm0.004}}             
\newcommand{\hatcurLBiz}{\ensuremath{0.1003}}                          
\newcommand{\hatcurLBiiz}{\ensuremath{0.3711}}                         
\newcommand{\hatcurLBir}{\ensuremath{0.2172}}                          
\newcommand{\hatcurLBiir}{\ensuremath{0.3951}}                         
\newcommand{\hatcurISOmshort}{\ensuremath{1.54}}                       
\newcommand{\hatcurISOmlong}{\ensuremath{1.543\pm0.051}}               
\newcommand{\hatcurISOrshort}{\ensuremath{1.83}}                       
\newcommand{\hatcurISOrlong}{\ensuremath{1.833_{-0.076}^{+0.138}}}     
\newcommand{\hatcurISOlogg}{\ensuremath{4.10\pm0.04}}                  
\newcommand{\hatcurISOlum}{\ensuremath{6.52_{-0.58}^{+1.07}}}          
\newcommand{\hatcurISOmv}{\ensuremath{2.67\pm0.14}}                    
\newcommand{\hatcurISOage}{\ensuremath{1.5\pm0.2}}                     
\newcommand{\hatcurISOMK}{\ensuremath{1.82\pm0.13}}                    
\newcommand{\hatcurRVK}{\ensuremath{188.7\pm21.9}}                     
\newcommand{\hatcurRVjitterA}{\ensuremath{59.9\pm19.4}}                
\newcommand{\hatcurRVjitterAround}{\ensuremath{60\pm19}}                
\newcommand{\hatcurRVjitterB}{\ensuremath{102.4\pm42.7}}               
\newcommand{\hatcurRVjitterBround}{\ensuremath{102\pm43}}               
\newcommand{\hatcurPPi}{\ensuremath{86.2\pm1.7}}                       
\newcommand{\hatcurPPlogg}{\ensuremath{3.33\pm0.07}}                   
\newcommand{\hatcurPPar}{\ensuremath{5.13_{-0.30}^{+0.19}}}            
\newcommand{\hatcurPParel}{\ensuremath{0.0438\pm0.0005}}               
\newcommand{\hatcurPPrho}{\ensuremath{0.75\pm0.17}}                    
\newcommand{\hatcurPPmshort}{\ensuremath{1.73}}                        
\newcommand{\hatcurPPmlong}{\ensuremath{1.730\pm0.205}}                
\newcommand{\hatcurPPrshort}{\ensuremath{1.41}}                        
\newcommand{\hatcurPPrlong}{\ensuremath{1.413_{-0.077}^{+0.128}}}      
\newcommand{\hatcurPPmrcorr}{\ensuremath{0.12}}                        
\newcommand{\hatcurPPteff}{\ensuremath{2131_{-42}^{+69}}}              
\newcommand{\hatcurPPtheta}{\ensuremath{0.069\pm0.009}}                
\newcommand{\hatcurPPfluxavg}{\ensuremath{4.65_{-0.36}^{+0.65}}}       
\newcommand{\hatcurXAv}{\ensuremath{0.111\pm0.037}}                    
\newcommand{\hatcurXdistred}{\ensuremath{322_{-13}^{+24}}}             
\newcommand{\hatcurSMEiteffcirc}{\ensuremath{6668\pm52}}               
\newcommand{\hatcurSMEizfehcirc}{\ensuremath{0.00\pm0.08}}             
\newcommand{\hatcurSMEizfehshortcirc}{\ensuremath{0.00}}               
\newcommand{\hatcurSMEiloggcirc}{\ensuremath{3.95\pm0.10}}             
\newcommand{\hatcurSMEivsincirc}{\ensuremath{16.02\pm0.50}}            
\newcommand{\hatcurSMEivmaccirc}{\ensuremath{NULL}}                    
\newcommand{\hatcurSMEivmiccirc}{\ensuremath{NULL}}                    
\newcommand{\hatcurSMEiiteffcirc}{\ensuremath{6820\pm52}}              
\newcommand{\hatcurSMEiizfehcirc}{\ensuremath{0.074\pm0.080}}          
\newcommand{\hatcurSMEiizfehshortcirc}{\ensuremath{0.074}}             
\newcommand{\hatcurSMEiiloggcirc}{\ensuremath{4.10\pm0.06}}            
\newcommand{\hatcurSMEiivsincirc}{\ensuremath{16.00\pm0.50}}           
\newcommand{\hatcurSMEiivmaccirc}{\ensuremath{NULL}}                   
\newcommand{\hatcurSMEiivmiccirc}{\ensuremath{NULL}}                   
\newcommand{\hatcurLCrprstareccen}{\ensuremath{0.0789\pm0.0019}}       
\newcommand{\hatcurLCbsqeccen}{\ensuremath{0.106_{-0.062}^{+0.102}}}   
\newcommand{\hatcurLCimpeccen}{\ensuremath{0.325_{-0.135}^{+0.122}}}   
\newcommand{\hatcurLCzetaeccen}{\ensuremath{12.75\pm0.10}}             
\newcommand{\hatcurLCdureccen}{\ensuremath{0.1708\pm0.0019}}           
\newcommand{\hatcurLCingdureccen}{\ensuremath{0.0138\pm0.0017}}        
\newcommand{\hatcurLCPeccen}{\ensuremath{2.691548\pm0.000007}}         
\newcommand{\hatcurLCTeccen}{\ensuremath{2456399.62418\pm0.00065}}     
\newcommand{\hatcurSMEiteffeccen}{\ensuremath{6668\pm52}}              
\newcommand{\hatcurSMEizfeheccen}{\ensuremath{0.00\pm0.08}}            
\newcommand{\hatcurSMEizfehshorteccen}{\ensuremath{0.00}}              
\newcommand{\hatcurSMEiloggeccen}{\ensuremath{3.95\pm0.10}}            
\newcommand{\hatcurSMEivsineccen}{\ensuremath{16.02\pm0.50}}           
\newcommand{\hatcurSMEivmaceccen}{\ensuremath{NULL}}                   
\newcommand{\hatcurSMEivmiceccen}{\ensuremath{NULL}}                   
\newcommand{\hatcurSMEiiteffeccen}{\ensuremath{6820\pm52}}             
\newcommand{\hatcurSMEiizfeheccen}{\ensuremath{0.074\pm0.080}}         
\newcommand{\hatcurSMEiizfehshorteccen}{\ensuremath{0.074}}            
\newcommand{\hatcurSMEiiloggeccen}{\ensuremath{4.10\pm0.09}}           
\newcommand{\hatcurSMEiivsineccen}{\ensuremath{16.00\pm0.50}}          
\newcommand{\hatcurSMEiivmaceccen}{\ensuremath{NULL}}                  
\newcommand{\hatcurSMEiivmiceccen}{\ensuremath{NULL}}                  
\newcommand{\hatcurLBizeccen}{\ensuremath{0.1003}}                     
\newcommand{\hatcurLBiizeccen}{\ensuremath{0.3711}}                    
\newcommand{\hatcurLBireccen}{\ensuremath{0.2172}}                     
\newcommand{\hatcurLBiireccen}{\ensuremath{0.3951}}                    
\newcommand{\hatcurISOmlongeccen}{\ensuremath{1.543\pm0.073}}          
\newcommand{\hatcurISOrlongeccen}{\ensuremath{1.838_{-0.169}^{+0.237}}} 
\newcommand{\hatcurISOloggeccen}{\ensuremath{4.10\pm0.08}}             
\newcommand{\hatcurISOlumeccen}{\ensuremath{6.55_{-1.14}^{+1.90}}}     
\newcommand{\hatcurISOmveccen}{\ensuremath{2.66\pm0.24}}               
\newcommand{\hatcurISOageeccen}{\ensuremath{1.4_{-0.3}^{+0.2}}}        
\newcommand{\hatcurISOMKeccen}{\ensuremath{1.82\pm0.24}}               
\newcommand{\hatcurRVKeccen}{\ensuremath{195.6\pm24.0}}                
\newcommand{\hatcurRVrkeccen}{\ensuremath{0.180_{-0.179}^{+0.120}}}    
\newcommand{\hatcurRVrheccen}{\ensuremath{0.002\pm0.216}}              
\newcommand{\hatcurRVkeccen}{\ensuremath{0.049_{-0.050}^{+0.065}}}     
\newcommand{\hatcurRVheccen}{\ensuremath{0.000\pm0.082}}               
\newcommand{\hatcurRVjitterAeccen}{\ensuremath{64.3\pm18.8}}           
\newcommand{\hatcurRVjitterAeccenround}{\ensuremath{64\pm19}}           
\newcommand{\hatcurRVjitterBeccen}{\ensuremath{91.4\pm40.3}}           
\newcommand{\hatcurRVjitterBeccenround}{\ensuremath{91\pm40}}           
\newcommand{\hatcurRVecceneccen}{\ensuremath{0.086\pm0.064}}           
\newcommand{\hatcurRVomegaeccen}{\ensuremath{173\pm132}}               
\newcommand{\hatcurPPieccen}{\ensuremath{86.4_{-2.1}^{+1.6}}}          
\newcommand{\hatcurPPloggeccen}{\ensuremath{3.34\pm0.10}}              
\newcommand{\hatcurPPareccen}{\ensuremath{5.12\pm0.48}}                
\newcommand{\hatcurPPareleccen}{\ensuremath{0.0438\pm0.0007}}          
\newcommand{\hatcurPPrhoeccen}{\ensuremath{0.78_{-0.21}^{+0.35}}}      
\newcommand{\hatcurPPmlongeccen}{\ensuremath{1.785\pm0.222}}           
\newcommand{\hatcurPPrlongeccen}{\ensuremath{1.411_{-0.138}^{+0.197}}} 
\newcommand{\hatcurPPmrcorreccen}{\ensuremath{0.16}}                   
\newcommand{\hatcurPPteffeccen}{\ensuremath{2134_{-90}^{+117}}}        
\newcommand{\hatcurPPthetaeccen}{\ensuremath{0.071_{-0.011}^{+0.014}}} 
\newcommand{\hatcurPPfluxavgeccen}{\ensuremath{4.69_{-0.72}^{+1.19}}}  
\newcommand{\hatcurXAveccen}{\ensuremath{0.110\pm0.037}}               
\newcommand{\hatcurXdistredeccen}{\ensuremath{323_{-29}^{+41}}}        

\newcommand{\hatcur}{HAT-P-49}
\newcommand{\hatcurb}{HAT-P-49b}

\newcommand{\hatcurisoshort}{YY}

\newcommand{\hatcurlumind}{\arstar}
\newcommand{\hatcurjhkfilset}{ESO}

\newcommand{\hatcurSMEversion}{ii}                                       
\newcommand{\hatcurSMEteff}{\ifthenelse{\equal{\hatcurSMEversion}{i}}{\hatcurSMEiteff}{\hatcurSMEiiteff}}
\newcommand{\hatcurSMEzfeh}{\ifthenelse{\equal{\hatcurSMEversion}{i}}{\hatcurSMEizfeh}{\hatcurSMEiizfeh}}
\newcommand{\hatcurSMEzfehshort}{\ifthenelse{\equal{\hatcurSMEversion}{i}}{\hatcurSMEizfehshort}{\hatcurSMEiizfehshort}}
\newcommand{\hatcurSMElogg}{\ifthenelse{\equal{\hatcurSMEversion}{i}}{\hatcurSMEilogg}{\hatcurSMEiilogg}}
\newcommand{\hatcurSMEvsin}{\ifthenelse{\equal{\hatcurSMEversion}{i}}{\hatcurSMEivsin}{\hatcurSMEiivsin}}
\newcommand{\hatcurSMEvmac}{\ifthenelse{\equal{\hatcurSMEversion}{i}}{\hatcurSMEivmac}{\hatcurSMEiivmac}}
\newcommand{\hatcurSMEvmic}{\ifthenelse{\equal{\hatcurSMEversion}{i}}{\hatcurSMEivmic}{\hatcurSMEiivmic}}

\newcommand{\hatcurSMEteffcirc}{\ifthenelse{\equal{\hatcurSMEversion}{i}}{\hatcurSMEiteffcirc}{\hatcurSMEiiteffcirc}}
\newcommand{\hatcurSMEzfehcirc}{\ifthenelse{\equal{\hatcurSMEversion}{i}}{\hatcurSMEizfehcirc}{\hatcurSMEiizfehcirc}}
\newcommand{\hatcurSMEzfehshortcirc}{\ifthenelse{\equal{\hatcurSMEversion}{i}}{\hatcurSMEizfehshortcirc}{\hatcurSMEiizfehshortcirc}}
\newcommand{\hatcurSMEloggcirc}{\ifthenelse{\equal{\hatcurSMEversion}{i}}{\hatcurSMEiloggcirc}{\hatcurSMEiiloggcirc}}
\newcommand{\hatcurSMEvsincirc}{\ifthenelse{\equal{\hatcurSMEversion}{i}}{\hatcurSMEivsincirc}{\hatcurSMEiivsincirc}}
\newcommand{\hatcurSMEvmaccirc}{\ifthenelse{\equal{\hatcurSMEversion}{i}}{\hatcurSMEivmaccirc}{\hatcurSMEiivmaccirc}}
\newcommand{\hatcurSMEvmiccirc}{\ifthenelse{\equal{\hatcurSMEversion}{i}}{\hatcurSMEivmiccirc}{\hatcurSMEiivmiccirc}}

\newcommand{\hatcurSMEteffeccen}{\ifthenelse{\equal{\hatcurSMEversion}{i}}{\hatcurSMEiteffeccen}{\hatcurSMEiiteffeccen}}
\newcommand{\hatcurSMEzfeheccen}{\ifthenelse{\equal{\hatcurSMEversion}{i}}{\hatcurSMEizfeheccen}{\hatcurSMEiizfeheccen}}
\newcommand{\hatcurSMEzfehshorteccen}{\ifthenelse{\equal{\hatcurSMEversion}{i}}{\hatcurSMEizfehshorteccen}{\hatcurSMEiizfehshorteccen}}
\newcommand{\hatcurSMEloggeccen}{\ifthenelse{\equal{\hatcurSMEversion}{i}}{\hatcurSMEiloggeccen}{\hatcurSMEiiloggeccen}}
\newcommand{\hatcurSMEvsineccen}{\ifthenelse{\equal{\hatcurSMEversion}{i}}{\hatcurSMEivsineccen}{\hatcurSMEiivsineccen}}
\newcommand{\hatcurSMEvmaceccen}{\ifthenelse{\equal{\hatcurSMEversion}{i}}{\hatcurSMEivmaceccen}{\hatcurSMEiivmaceccen}}
\newcommand{\hatcurSMEvmiceccen}{\ifthenelse{\equal{\hatcurSMEversion}{i}}{\hatcurSMEivmiceccen}{\hatcurSMEiivmiceccen}}

\newcounter{planetcounter}

\newboolean{emulateapj}
\setboolean{emulateapj}{true}

\newboolean{rvtablelong}
\setboolean{rvtablelong}{true}

\newboolean{astroph}
\setboolean{astroph}{true}

\shortauthors{Bieryla et al.}
\shorttitle{
\hatcur\lowercase{b}
}
\ifthenelse{\boolean{emulateapj}}{
    \newcommand{\titledag}{$\dagger$}
}{
    \newcommand{\titledag}{\dagger}
}

\begin{document}
\title{
\hatcur\lowercase{b}: A 1.7\,\mjup\ Planet Transiting a Bright 1.5\,\msun\ F-Star \altaffilmark{\titledag}
}

\author{
    A.~Bieryla\altaffilmark{1},
    J.~D.~Hartman\altaffilmark{2},
    G.~\'A.~Bakos\altaffilmark{2,7},
    W.~Bhatti\altaffilmark{2}, 
    G.~Kov\'acs\altaffilmark{3,6}, 
    I.~Boisse\altaffilmark{4},
    D.~W.~Latham\altaffilmark{1},
    L.~A.~Buchhave\altaffilmark{1,5},
    Z.~Csubry\altaffilmark{2},
    K.~Penev\altaffilmark{2},
    M.~de Val-Borro\altaffilmark{2},
    B.~B\'eky\altaffilmark{1},
    E.~Falco\altaffilmark{1},
    G.~Torres\altaffilmark{1},
    R.~W.~Noyes\altaffilmark{1},
    P.~Berlind\altaffilmark{1},
    M.~C.~Calkins\altaffilmark{1},
    G.~A.~Esquerdo\altaffilmark{1},
    J.~L\'az\'ar\altaffilmark{8},
    I.~Papp\altaffilmark{8},
    P.~S\'ari\altaffilmark{8}
}

\altaffiltext{1}{Harvard-Smithsonian Center for Astrophysics,
    Cambridge, MA 02138 USA; email: abieryla@cfa.harvard.edu}

\altaffiltext{2}{Department of Astrophysical Sciences, Princeton
  University, Princeton, NJ 08544 USA; email: gbakos@astro.princeton.edu}

\altaffiltext{3}{Konkoly Observatory, Budapest, Hungary}

\altaffiltext{4}{Centro de Astrof\'isica, Universidade do Porto, Rua das Estrelas, 4150-762 Porto, Portugal}

\altaffiltext{5}{Niels Bohr Institute, University of Copenhagen, DK-2100, Denmark, and Centre for Star and Planet Formation, Natural History Museum of Denmark, DK-1350 Copenhagen}

\altaffiltext{6}{Department of Physics and Astrophysics, University of North Dakota, Grand Forks, ND USA}

\altaffiltext{7}{Sloan Fellow}

\altaffiltext{8}{Hungarian Astronomical Association (HAA).}

\altaffiltext{$\dagger$}{
    Based on observations obtained with the Hungarian-made Automated
    Telescope Network. Based in part on observations obtained with the
    Tillinghast Reflector 1.5\,m telescope and the 1.2\,m telescope,
    both operated by the Smithsonian Astrophysical Observatory at the
    Fred Lawrence Whipple Observatory in Arizona. Based in part on radial
    velocities obtained with the SOPHIE spectrograph mounted on the
    1.93\,m telescope at Observatoire de Haute Provence, France.
}


\begin{abstract}

\setcounter{footnote}{10}
We report the discovery of the transiting extrasolar planet \hatcurb{}. 
The planet transits the bright ($V = \hatcurCCtassmvshort$) slightly evolved 
F-star \hatcurCChd\ with a mass of $\hatcurISOmshort$\,\msun\ and a radius of
$\hatcurISOrshort$\, \rsun.  \hatcurb{} is orbiting one of the $25$ brightest 
stars to host a transiting planet which makes this a favorable candidate
for detailed follow-up.  
This system is an especially strong target for Rossiter-McLaughlin follow-up 
due to the host star's fast rotation, $16$ {\kms}.
The planetary companion has a period of 
$\hatcurLCPshort$\,d, mass of $\hatcurPPmshort$\,\mjup\ and radius 
of $\hatcurPPrshort$\,\rjup. The planetary characteristics are consistent 
with that of a classical hot Jupiter but we note that this is the fourth 
most massive star to host a transiting planet with both \mpl\ and \rpl\
well determined. 

\setcounter{footnote}{0}
\end{abstract}



\section{Introduction}
\label{sec:introduction}

We are in an exciting time for exoplanet discovery. There are over 900 
confirmed exoplanets, with a third of those planets transiting their host star.
Transiting extrasolar planets (TEPs) are important because they provide information 
about the planet's mass and radius and further follow-up can be obtained  
to study the spin-orbit alignment of the star and even impart details about the 
atmospheric composition of the planet. Planets with bright host stars are 
appealing to study because they are targets that can provide precise information 
using less telescope time. Of the known transiting planets, there are only a couple dozen planets 
that transit host stars bright enough to do detailed follow-up studies and even fewer 
planets that orbit so close to their massive host stars.  

In \refsecl{obs} we summarize the detection of the photometric transit
signal and the subsequent spectroscopic and photometric observations
of the star to confirm the planet. In \refsecl{analysis} we analyze
the data to rule out false positive scenarios, and to determine the
stellar and planetary parameters. Our findings are briefly discussed in
\refsecl{discussion}.

\section{Observations}
\label{sec:obs}

The general procedure used by HATNet to discover TEPs is described in previous papers
\citep[e.g.][]{bakos:2010:hat11,latham:2009:hat8}. In this section, we describe 
the initial discovery of \hatcurb{}, a new transiting extrasolar planet, found by the Hungarian-made 
Automated Telescope Network \citep[HATNet;][]{bakos:2004:hatnet} survey. The new planet 
orbiting \hatcurCChd\ was confirmed with photometry using KeplerCam on the 1.2 m telescope 
at the Fred Lawrence Whipple Observatory (FLWO) in Arizona and with spectroscopy using the 
Tillinghast Reflector Echelle Spectrograph (TRES) on the 1.5 m telescope at FLWO and the SOPHIE spectrograph 
on the Observatoire de Haute Provence (OHP) 1.93 m telescope.
Identifying information for this star is provided in \reftabl{stellar}.

\subsection{Photometric detection}
\label{sec:detection}

\begin{figure}[]
\plotone{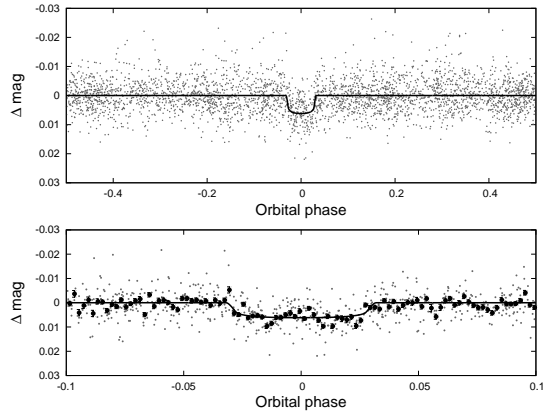}
\caption[]{
    HATNet \lc{} of \hatcur\ phase folded with the transit period.
    The top panel shows the unbinned light curve,
    while the bottom shows the region zoomed-in on the transit, with
    dark filled circles for the light curve binned in phase with a
    binsize of 0.002. The solid line shows the model fit to the light
    curve.
\label{fig:hatnet}}
\end{figure}

The initial identification of \hatcur{} as a potential transiting
planet system was based on photometric observations made with the
fully automated HATNet system. A $10.6\degr \times 10.6\degr$ field
containing \hatcur\ was observed between 15 September 2008 and 19 May
2009 using the HAT-7 telescope at FLWO in Arizona, and the HAT-8 telescope
at Mauna Kea Observatory in Hawaii. A total of 632 and 2589 images
containing \hatcur\ were obtained with HAT-7 and HAT-8,
respectively\footnote{These numbers exclude images whose photometry produced
outliers that were culled from the resulting light
curves.}. Observations were made through a Sloan~\band{r} filter using
$300$\,s exposures with a median cadence of $360$\,s. The data were
reduced to noise-filtered light curves following
\cite{bakos:2010:hat11}. These \lcs{} were searched for periodic
transit signals using the Box Least-Squares \citep[BLS;
  see][]{kovacs:2002:BLS} method, leading to the identification of
\hatcur\ as a candidate TEP system with a transit \lc{} depth of $\hatcurLCdip{}$
mmag and a period of $\hatcurLCPshort{}$ days (\reffigl{hatnet}). The 
trend-filtered differential photometric measurements for \hatcur{} are
given in Table~\ref{tab:phfu}.
We subtracted the transit signal from the HAT light curve and used BLS to search for additional transit signals at other periods. We find no other significant transit signals with a depth greater than 10\,mmag and a frequency between 0 and 2.0\,d$^{-1}$. We also searched the light curve for periodic variability (e.g.~due to pulsations or spots) using the Discrete Fourier Transform and do not find any signals with amplitudes above 0.6\,mmag in the 0.0 to 50.0\,d$^{-1}$ frequency range.

\subsection{Spectroscopy}
\label{sec:hispec}

We proceeded with the follow-up by obtaining spectra to rule out false
positives, to characterize the RV variations, and to refine the
determination of the stellar parameters. Spectroscopic observations
were made with the Tillinghast Reflector Echelle Spectrograph
\citep[TRES;][]{furesz:2008} on the Tillinghast Reflector 1.5\,m telescope at
FLWO in Arizona, and with the SOPHIE spectrograph \citep{bouchy:2009} on the
Observatoire de Haute Provence (OHP) 1.93\,m telescope. We obtained
18 spectra with TRES and 6 spectra with SOPHIE. 

The first five TRES spectra were obtained with exposure times of $180$
to $450$\,s yielding a S/N calculated near the MgB region ranging from $28$ to
$47$. We derived initial RV measurements and stellar atmospheric
parameters (including the effective surface temperature $\teff$, the
surface gravity $\logg$ and the projected equatorial rotation speed
$\vsini$) following the method of
\cite{buchhave:2010:hat16}. 
These results allowed us to confirm that the host star is on the main sequence, that it does not have an obviously composite spectrum, and that it does not show the large velocity variations that would be present if this were an F-M binary.
We then proceeded by obtaining
additional higher S/N observations using both TRES and SOPHIE.

The Stellar Parameter Classification (SPC) fitting program
\citep{buchhave:2012} was used to determine the final spectroscopic 
parameters of the host star. The values were calculated using a
weighted mean, taking into account the cross correlation function (CCF) peak height. 
The results from this analysis are shown in Table 3.  

A multi-order velocity analysis was done using the TRES spectra. Each
observed spectrum was cross-correlated, order by order, using the 
strongest observed spectrum as a template. Fourteen orders were used
excluding the bluest orders due to low S/N per resolution element, the reddest 
orders due to fringing and a few orders with known telluric absorption lines.

\begin{figure} []
\plotone{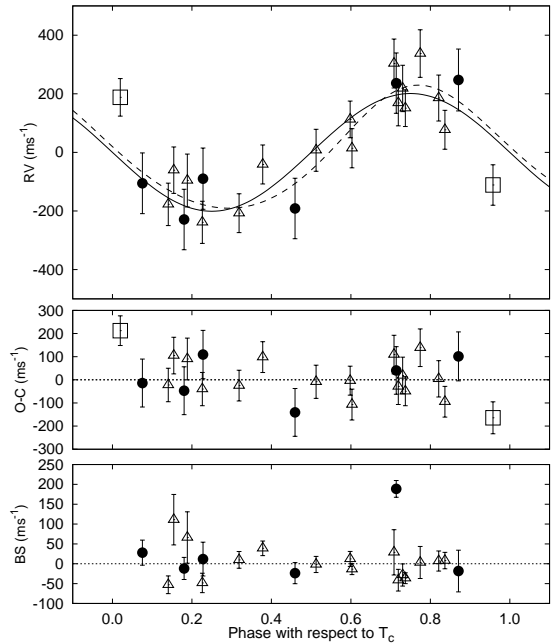}
\caption{
    {\em Top panel:} RV measurements from FLWO~1.5\,m/TRES (open
    triangles) and OHP~1.93\,m/SOPHIE (filled circles) for
    \hbox{\hatcur} shown as a function of orbital phase, along with
    our best-fit circular model (solid line; see
    \reftabl{planetparam}), and our best-fit eccentric model (dashed
    line).  Zero phase corresponds to the time of mid-transit.  The
    center-of-mass velocity has been subtracted. Two open squares are
    RV measurements from FLWO~1.5\,m/TRES that were obtained during
    transit and have been excluded from the analysis due to the
    possibility that they are affected by the Rossiter-McLaughlin
    effect which is not included in our model.
    {\em Second panel:} Velocity $O\!-\!C$ residuals from the best
    fit. The error bars include a ``jitter'' component
    (\hatcurRVjitterAround\,\ms, and \hatcurRVjitterBround\,\ms\ for TRES and
    SOPHIE respectively) added in quadrature to the formal errors (see
    \refsecl{hispec}). The symbols are as in the upper panel.
    {\em Third panel:} Bisector spans (BS), with the mean value
    subtracted. The measurement from the template spectrum is
    included. BS uncertainties were estimated using the relation
    $\sigma_{BS} = 2\sigma_{RV}$. 
    Note the different vertical scales of the panels.
}
\label{fig:rvbis}
\end{figure}

Observations with SOPHIE were carried out, and reduced to radial
velocities and spectral line bisector spans (BSs) measurements following
\citet{boisse:2012:hat4243}. 
The radial velocities and BSs measured from 
these spectra are provided in \reftabl{rvs} and the
computed BSs are also shown in ~\reffigl{rvbis}. The BSs allowed us to rule out
various blend possibilities because there was no significant variation
in phase with the RVs.

\setcounter{planetcounter}{1}
%

\subsection{Photometric follow-up observations}
\label{sec:phot}
 
\begin{figure}[]
\plotone{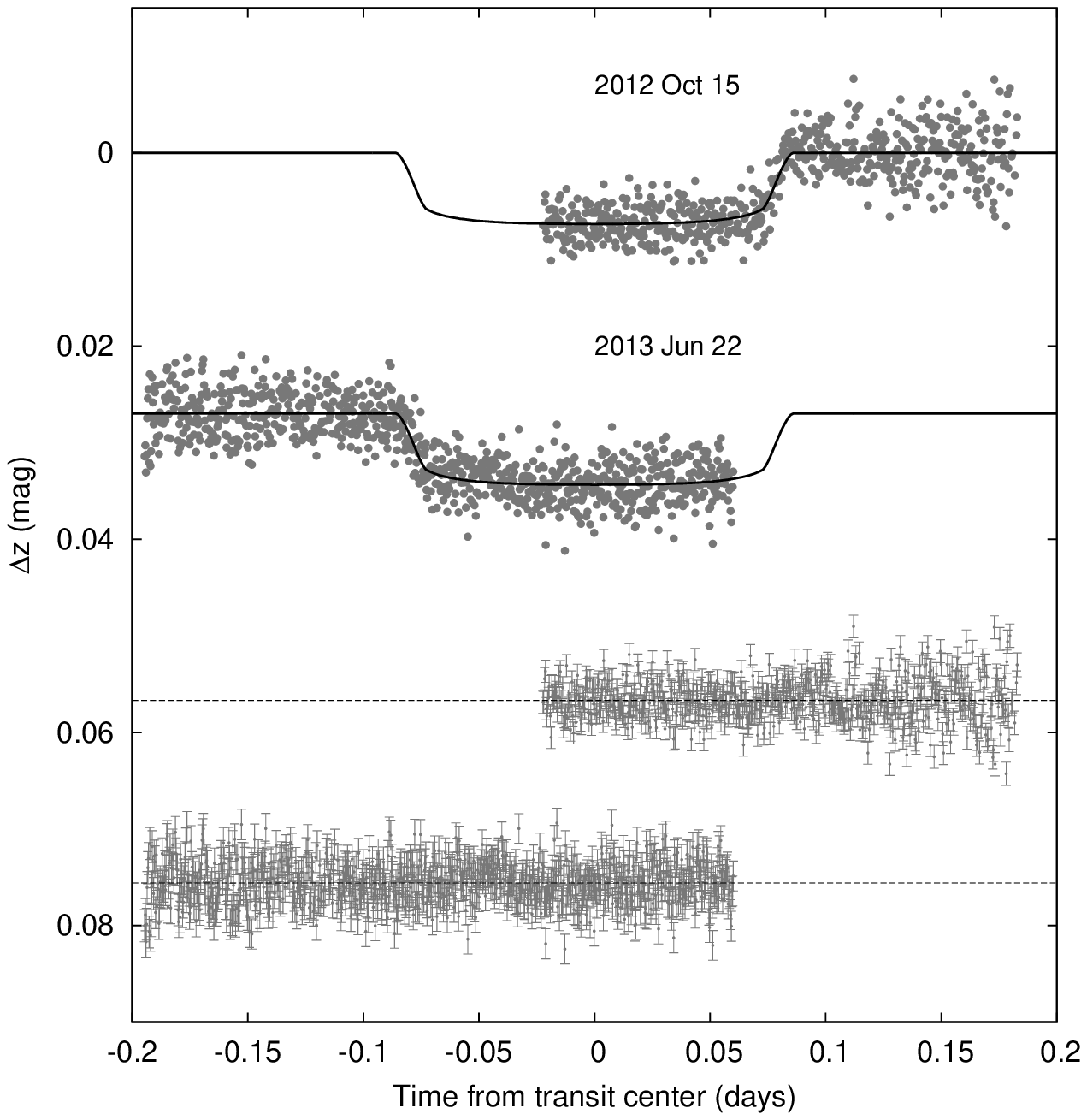}
\caption{
    Unbinned transit \lcs{} for \hatcur, acquired with KeplerCam at
    the \flwof{} telescope.  The light curves have been EPD- and
    TFA-processed, as described in \citet{bakos:2010:hat11}.  The
    dates of each event are indicated. Our best fit from the global
    modeling described in \refsecl{analysis} is shown by the solid
    line.  Residuals from the fit are displayed below in the same
    order as the original light curves.  The error bars represent the
    photon and background shot noise, plus the readout noise.
}
\label{fig:lc}
\end{figure}

We conducted additional photometric observations of \hatcur\ using
KeplerCam on the \flwof{} telescope. We observed a transit
egress in the Sloan \band{z} on the night of 15 October 2012, and an ingress
in the Sloan \band{z} on the night of 22 June 2013. For the first event we
obtained 603 images with an exposure time of 15\,s and a median
cadence of $29$\,s. For the second event we obtained 875 images with
an exposure time of 10\,s and a median cadence of $24.2$\,s. The
images were reduced to light curves following
\citet{bakos:2010:hat11}. We performed External Parameter
Decorrelation \citep[EPD; see][]{bakos:2010:hat11} and Trend Filtering 
Algorithm \citep[TFA; see][]{kovacs:2005:TFA} to remove trends
simultaneously with the light curve modeling. 
The final time series, together with our
best-fit transit \lc{} models, are shown in the top portion of
Figure~\ref{fig:lc}, while the individual measurements are reported in
Table~\ref{tab:phfu}.

\ifthenelse{\boolean{emulateapj}}{
    \begin{deluxetable*}{lrrrrr}
}{
    \begin{deluxetable}{lrrrrr}
}
\tablewidth{0pc}
\tablecaption{
    Differential photometry of
    \hatcur\label{tab:phfu}.
}
\tablehead{
    \colhead{BJD\tablenotemark{a}} & 
    \colhead{Mag\tablenotemark{b}} & 
    \colhead{\ensuremath{\sigma_{\rm Mag}}} &
    \colhead{Mag(orig)\tablenotemark{c}} & 
    \colhead{Filter} &
    \colhead{Instrument} \\
    \colhead{\hbox{~~~~(2,400,000$+$)~~~~}} & 
    \colhead{} & 
    \colhead{} &
    \colhead{} & 
    \colhead{} & 
    \colhead{}
}
\startdata
$ 54769.89219 $ & $  -0.00083 $ & $   0.00256 $ & $ \cdots $ & $ r$ &     HATNet\\
$ 54753.74292 $ & $   0.00389 $ & $   0.00241 $ & $ \cdots $ & $ r$ &     HATNet\\
$ 54796.80816 $ & $   0.00120 $ & $   0.00267 $ & $ \cdots $ & $ r$ &     HATNet\\
$ 54726.82816 $ & $   0.00655 $ & $   0.00313 $ & $ \cdots $ & $ r$ &     HATNet\\
$ 54761.81965 $ & $   0.00371 $ & $   0.00213 $ & $ \cdots $ & $ r$ &     HATNet\\
$ 54734.90461 $ & $   0.01027 $ & $   0.00218 $ & $ \cdots $ & $ r$ &     HATNet\\
$ 54753.74710 $ & $  -0.00165 $ & $   0.00224 $ & $ \cdots $ & $ r$ &     HATNet\\
$ 54726.83223 $ & $   0.00083 $ & $   0.00259 $ & $ \cdots $ & $ r$ &     HATNet\\
$ 54734.90877 $ & $  -0.00272 $ & $   0.00220 $ & $ \cdots $ & $ r$ &     HATNet\\
$ 54761.82432 $ & $  -0.00367 $ & $   0.00204 $ & $ \cdots $ & $ r$ &     HATNet\\

\enddata
\tablenotetext{a}{
    Barycentric Julian Date calculated directly from UTC, {\em
      without} correction for leap seconds.
}
\tablenotetext{b}{
    The out-of-transit level has been subtracted. These magnitudes have
    been subjected to the EPD and TFA procedures, carried out
    simultaneously with the transit fit.
}
\tablenotetext{c}{
    Raw magnitude values after correction using comparison stars, but
    without application of the EPD and TFA procedures. This is only
    reported for the follow-up light curves.
}
\tablecomments{
    This table is available in a machine-readable form in the online
    journal.  A portion is shown here for guidance regarding its form
    and content.
}
\ifthenelse{\boolean{emulateapj}}{
    \end{deluxetable*}
}{
    \end{deluxetable}
}

\ifthenelse{\boolean{emulateapj}}{
    \begin{deluxetable*}{lrrrrrrrrrr}
}{
    \begin{deluxetable}{lrrrrrrrrrr}
}
\tablewidth{0pc}
\tablecaption{
    Relative radial velocities, bisector span measurements and stellar
    atmospheric parameters of \hatcur.
    \label{tab:rvs}
}
\tablehead{
    \colhead{BJD\tablenotemark{a}} &
    \colhead{RV\tablenotemark{b}} &
    \colhead{\ensuremath{\sigma_{\rm RV}}\tablenotemark{c}} &
    \colhead{BS} &
    \colhead{\ensuremath{\sigma_{\rm BS}}} &
    \colhead{SNRe\tablenotemark{d}} &
    \colhead{\teff\tablenotemark{e}} &
    \colhead{\feh\tablenotemark{e}} &
    \colhead{\vsini\tablenotemark{e}} &
    \colhead{Phase} &
    \colhead{Instrument}\\
    \colhead{\hbox{(2,454,000$+$)}} &
    \colhead{(\ms)} &
    \colhead{(\ms)} &
    \colhead{(\ms)} &
    \colhead{(\ms)} &
    \colhead{} &
    \colhead{(K)} &
    \colhead{} &
    \colhead{(\kms)} &
    \colhead{} &
    \colhead{}
}
\startdata
$ 2110.90255 $ & $   218 $ & $    52 $ & $  -28 $ & $    28 $  & $ 47.6 $ & $ 6836 $ & $ 0.06 $ & $ 16.3 $ & $   0.730 $ & TRES \\
$ 2176.73278 $ & $   -96 $ & $    68 $ & $   66 $ & $    65 $  & $ 28.2 $ & $ 6594 $ & $ 0.28 $ & $ 17.3 $ & $   0.188 $ & TRES \\
$ 2192.79011 $ & $   -61 $ & $    52 $ & $  111 $ & $    64 $  & $ 30.5 $ & $ 6750 $ & $ 0.24 $ & $ 16.2 $ & $   0.154 $ & TRES \\
$ 2197.41035 $ & $   247 $ & $    26 $ & $  -18 $ & $    53 $ & \nodata      & \nodata      & \nodata      & \nodata      & $   0.871 $ & SOPHIE \\
$ 2198.37133 $ & $   -90 $ & $    21 $ & $   12 $ & $    43 $ & \nodata      & \nodata      & \nodata      & \nodata      & $   0.228 $ & SOPHIE \\
$ 2199.66505 $ & $   303 $ & $    58 $ & $   29 $ & $    57 $ & $ 33.2 $ & $ 6768 $ & $ 0.21 $ & $ 16.5 $ & $   0.708 $ & TRES \\
$ 2202.37199 $ & $   236 $ & $    11 $ & $  189 $ & $    21 $ & \nodata      & \nodata      & \nodata      & \nodata      & $   0.714 $ & SOPHIE \\
$ 2203.34402 $ & $  -106 $ & $    16 $ & $   28 $ & $    32 $ & \nodata      & \nodata      & \nodata      & \nodata      & $   0.075 $ & SOPHIE \\
$ 2204.37766 $ & $  -192 $ & $    13 $ & $  -23 $ & $    27 $ & \nodata      & \nodata      & \nodata      & \nodata      & $   0.459 $ & SOPHIE \\
$ 2206.31791 $ & $  -229 $ & $    14 $ & $  -12 $ & $    28 $ & \nodata      & \nodata      & \nodata      & \nodata      & $   0.180 $ & SOPHIE \\
$ 2210.60885 $ & $   337 $ & $    55 $ & $    3 $ & $    40 $ & $ 37.4 $ & $ 6898 $ & $ 0.16 $ & $ 16.5 $ & $   0.774 $ & TRES \\
$ 2223.59189 $ & $   113 $ & $    19 $ & $   12 $ & $    19 $ & $ 89.2 $ & $ 6816 $ & $ 0.00 $ & $ 15.7 $ & $   0.598 $ & TRES \\
$ 2224.72636 $\tablenotemark{f} & $   188 $ & $    23 $ & $   -32 $ & $   25 $ & $ 85.7 $ & $ 6824 $ & $ 0.03 $ & $ 15.8 $ & $   0.020 $ & TRES \\
$ 2225.69126 $ & $   -41 $ & $    29 $ & $   39 $ & $    18 $ & $ 79.3 $ & $ 6826 $ & $ 0.03 $ & $ 15.9 $ & $   0.378 $ & TRES \\
$ 2226.65734 $ & $   151 $ & $    19 $ & $  -36 $ & $    14 $ & $ 92.3 $ & $ 6824 $ & $ 0.03 $ & $ 15.7 $ & $   0.737 $ & TRES \\
$ 2227.74319 $ & $  -177 $ & $    41 $ & $  -53 $ & $    22 $ & $ 75.8 $ & $ 6809 $ & $ 0.03 $ & $ 15.8 $ & $   0.140 $ & TRES \\
$ 2228.74445 $ & $     7 $ & $    39 $ & $   -2 $ & $    20 $ & $ 80.5 $ & $ 6825 $ & $ 0.03 $ & $ 15.7 $ & $   0.512 $ & TRES \\
$ 2229.61724 $ & $    77 $ & $    29 $ & $    8 $ & $    21 $ & $ 80.5 $ & $ 6829 $ & $ 0.04 $ & $ 15.9 $ & $   0.837 $ & TRES \\
$ 2230.66549 $ & $  -239 $ & $    40 $ & $  -48 $ & $    25 $ & $ 86.6 $ & $ 6840 $ & $ 0.04 $ & $ 15.9 $ & $   0.226 $ & TRES \\
$ 2231.67934 $ & $    14 $ & $    30 $ & $  -14 $ & $    14 $ & $ 85.5 $ & $ 6827 $ & $ 0.03 $ & $ 16.0 $ & $   0.603 $ & TRES \\
$ 2232.63450 $\tablenotemark{f} & $  -111 $ & $    34 $ & $   -30 $ & $   27 $ & $ 81.3 $ & $ 6827 $ & $ 0.04 $ & $ 15.9 $ & $   0.958 $ & TRES \\
$ 2233.60611 $ & $  -208 $ & $    29 $ & $   10 $ & $    21 $ & $ 83.4 $ & $ 6800 $ & $ 0.03 $ & $ 15.6 $ & $   0.319 $ & TRES \\
$ 2234.68416 $ & $   169 $ & $    50 $ & $  -41 $ & $    27 $ & $ 77.3 $ & $ 6819 $ & $ 0.01 $ & $ 15.9 $ & $   0.719 $ & TRES \\
$ 2237.64971 $ & $   186 $ & $    51 $ & $    7 $ & $    25 $ & $ 76.5 $ & $ 6844 $ & $ 0.04 $ & $ 15.8 $ & $   0.821 $ & TRES \\

\enddata
\tablenotetext{a}{
    Barycentric Julian Date calculated directly from UTC, {\em
      without} correction for leap seconds.
}
\tablenotetext{b}{
    The zero-point of these velocities is arbitrary. An overall offset
    $\gamma_{\rm rel}$ fitted to these velocities in \refsecl{analysis}
    has {\em not} been subtracted. 
}
\tablenotetext{c}{
    Internal errors excluding the component of astrophysical jitter
    considered in \refsecl{analysis}.
}
\tablenotetext{d}{
    Signal to noise per resolution element (SNRe) which takes into account the resolution of the instrument. SNRe is calculated near the MgB region.
}
\tablenotetext{e}{
    Spectroscopic parameters measured from the individual TRES spectra using SPC with the surface gravity fixed to $\loggstar = \hatcurISOlogg$ as determined from our global modelling. The uncertainties are $\sim 50$\,K, $0.08$\,dex and $0.5$\,\kms\ on $\teff$, $\feh$ and $\vsini$, respectively. We note that due to the rapid rotation of this star there is some discrepancy in the stellar classification of the observations with lower SNRe. The observations with lower SNRe have lower temperature and higher metallicity and we found that due to the rapid rotation of the star we needed a higher SNRe to get reliable classifications.
}
\tablenotetext{f}{
    These observations were obtained during transit and were excluded from our modelling of the orbit.
}
\ifthenelse{\boolean{rvtablelong}}{
}{
} 
\ifthenelse{\boolean{emulateapj}}{
    \end{deluxetable*}
}{
    \end{deluxetable}
}

\section{Analysis}
\label{sec:analysis}

\begin{figure}[]
\plotone{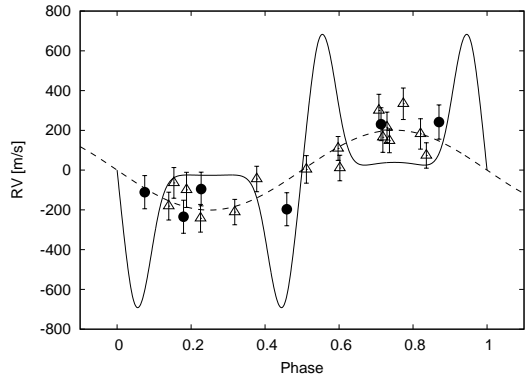}
\caption[]{
Comparison between the RV curve that would be measured from the cross-correlation function of a blended eclipsing binary system (solid line) and the observed RV measurements from TRES (open triangles) and SOPHIE (filled circles). The dashed line shows the best-fit orbit for a single-star plus transiting planet. The blend model shown here is for a case that cannot be rejected from the photometry alone, but can be rejected based on the RVs. This particular example, consisting of a $M = 1.64$\,\msun\ foreground star at a distance of 420\,pc blended with a background eclipsing binary with component masses of $1.8$\,\msun\ and $0.93$\,\msun\ and a distance of 2.2\,kpc, provides the best match to the RV data out of the blend models considered that fit the photometric data. Other blend models that fit the photometric data either predict much larger RV variations than observed (amplitudes above 1\,\kms), or they show non-sinusoidal variations that are similar to what is illustrated here.
\label{fig:blendrv}}
\end{figure}

To rule out blend scenarios that could potentially explain the
observations of \hatcur{} we conducted an analysis similar to that
done in \cite{hartman:2011:hat32hat33,hartman:2012:hat39hat41}. This
involved modeling the available light curves, absolute photometry, and
stellar atmospheric parameters as a combination of the observed primary 
and the postulated blended eclipsing binary using
the Padova isochrones \citep{girardi:2002} to constrain the properties
of the stars in the simulated systems. For each simulation we also
predicted the RVs and BS values that would have been measured from the
composite spectrum with FLWO~1.5\,m/TRES and OHP~1.93\,m/SOPHIE at the
times of observation. We found that although blend models covering a
relatively broad range of parameter space fit the photometry, most of
those that fit would result in several \kms\ RV variations that are
not observed, and all would result in RV variations with a scatter
that is at least a factor of 3 greater than what is
observed. Moreover, in those cases where predicted RVs vary by less
than 1 \kms, the variation is not sinusoidal in phase with the orbital
period (see \reffigl{blendrv}). We conclude that \hatcur{} is a transiting 
planet system and
not a blended stellar eclipsing binary system. The star-star-planet
degeneracy is generic in nearly all systems discovered so far, but none
of the subsequent detailed follow-up observations led to a different than
the original simplest star-planet model in any well-documented cases.
Therefore, we also accept the simplest model for this system as the most
plausible one.

We analyzed the system following
\cite{bakos:2010:hat11,hartman:2012:hat39hat41}. To summarize: (1) we
determined stellar atmospheric parameters for the host star by applying the
Stellar Parameter Classification method \citep{buchhave:2012} to the
FLWO~1.5\,m/TRES spectra; (2) we then conducted a Markov-Chain Monte
Carlo (MCMC) modeling of the available light curves and RVs fixing the
limb darkening coefficients to values computed by \cite{claret:2004}
for the measured atmospheric parameters; (3) we used the effective
temperatures and metallicities of the stars measured from the spectra,
together with the stellar densities determined from the joint light curve
and RV modeling to determine the stellar properties based on the Y$^2$
theoretical stellar evolution models \citep{yi:2001}. The stellar
properties so-determined include the masses, radii and ages. We also
determined the planetary parameters (e.g.,~mass and radius) which depend
on these values; (4) we re-analyzed the TRES spectra fixing the stellar
surface gravities to the values found in (3), and we then re-iterated steps
(2) and (3).

As is often the case, the photometric and RV observations show greater
scatter about the best-fit model than is expected based on the formal
uncertainties. To ensure that the resulting uncertainties on the
system parameters are not underestimated we inflated both the
photometric and RV uncertainties. The photometric uncertainties are
scaled by a factor such that $\chi^2 / {\rm d.o.f.} = 1$ for a given
light curve, for the best-fit model (the factors are $1.93$ and $1.44$
for the first and second KeplerCam light curves, respectively, and
$2.20$ for the HATNet light curve). We also added a ``jitter'' term in
quadrature to the RV uncertainties. The jitter term was allowed to vary
in the fit assuming a prior inversely proportional to the jitter. We did
this, rather than fixing the jitter, to allow a fair comparison
between different classes of models for the RV data, as discussed
below. We allowed independent jitters for each instrument as it is not
clear whether the jitter is instrumental or astrophysical in
origin. The median values (and $1\sigma$ uncertainties) for the
jitter, estimated from the parameter posterior distributions resulting
from our fit, are $\hatcurRVjitterAround{}$\,\ms\ and
$\hatcurRVjitterBround{}$\,\ms\ for the TRES and SOPHIE observations,
respectively, when the eccentricity is fixed to 0. When the
eccentricity is allowed to vary we find values of
$\hatcurRVjitterAeccenround{}$\,\ms\ and $\hatcurRVjitterBeccenround{}$\,\ms,
for TRES and SOPHIE, respectively.

We conducted the analysis twice: fixing the eccentricity to zero, and
allowing it to vary. We refer to these as the fixed-circular, and
free-eccentricity models, respectively.
We use the algorithm of \cite{weinberg:2013} to estimate the Bayesian
evidences for each model directly from the Markov Chains, and found
that the evidence ratio of the fixed-circular model to the
free-eccentricity model is $Z_{1}/Z_{2} \approx 10^6$. This indicates
that the data strongly favor the fixed-circular over the
free-eccentricity model, so we suggest adopting the system parameters
from the fixed-circular model.

The resulting derived stellar parameters and planetary parameters are
given in Tables~\ref{tab:stellar} and~\ref{tab:planetparam},
respectively, with the adopted parameters being those from the
fixed-circular model. We find that the star \hatcur{} has a mass of
$\hatcurISOmlong$\,\msun, and a radius of \hatcurISOrlong\,\rsun,
while the planet \hatcurb{} has a mass of $\hatcurPPmlong$\,\mjup, and
a radius of \hatcurPPrlong\,\rjup. The 95\% upper limit on the
eccentricity in the free-eccentricity model is $e < 0.212$.

\begin{figure}[]
\plotone{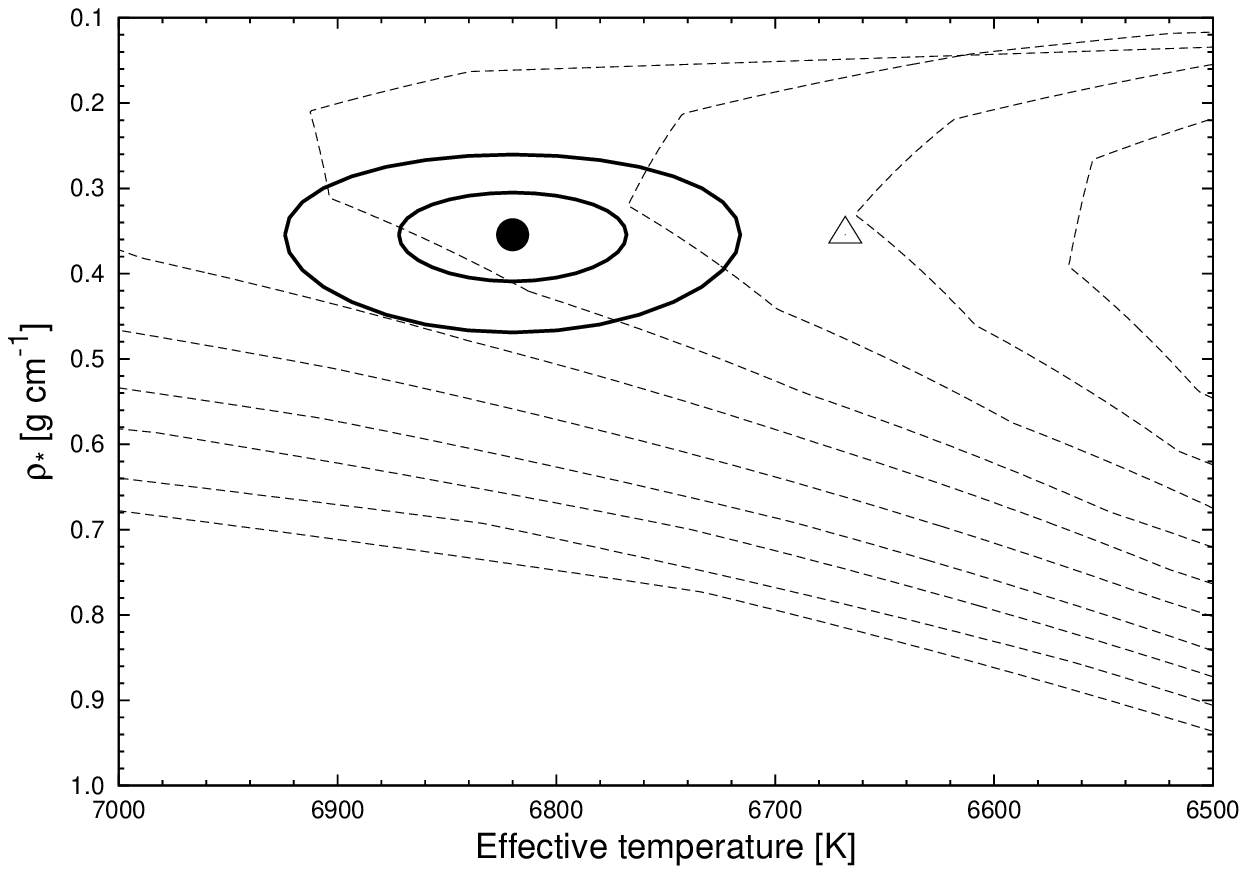}
\caption[]{
    Model isochrones from \citet{yi:2001} for the
    metallicity of \hatcur{}. The isochrones are shown for ages between 0.2\,Gyr and
    2.0\,Gyr in 0.2\,Gyr increments (age increases from left to
    right). The adopted
    values of $\teffstar$ and \rhostar\ are shown together with their
    1$\sigma$ and 2$\sigma$ confidence ellipsoids. The
    initial values of \teffstar\ and \rhostar\ from the first SPC and
    \lc\ analysis are represented with a triangle.
\label{fig:iso}}
\end{figure}
\begin{figure}[]
\plotone{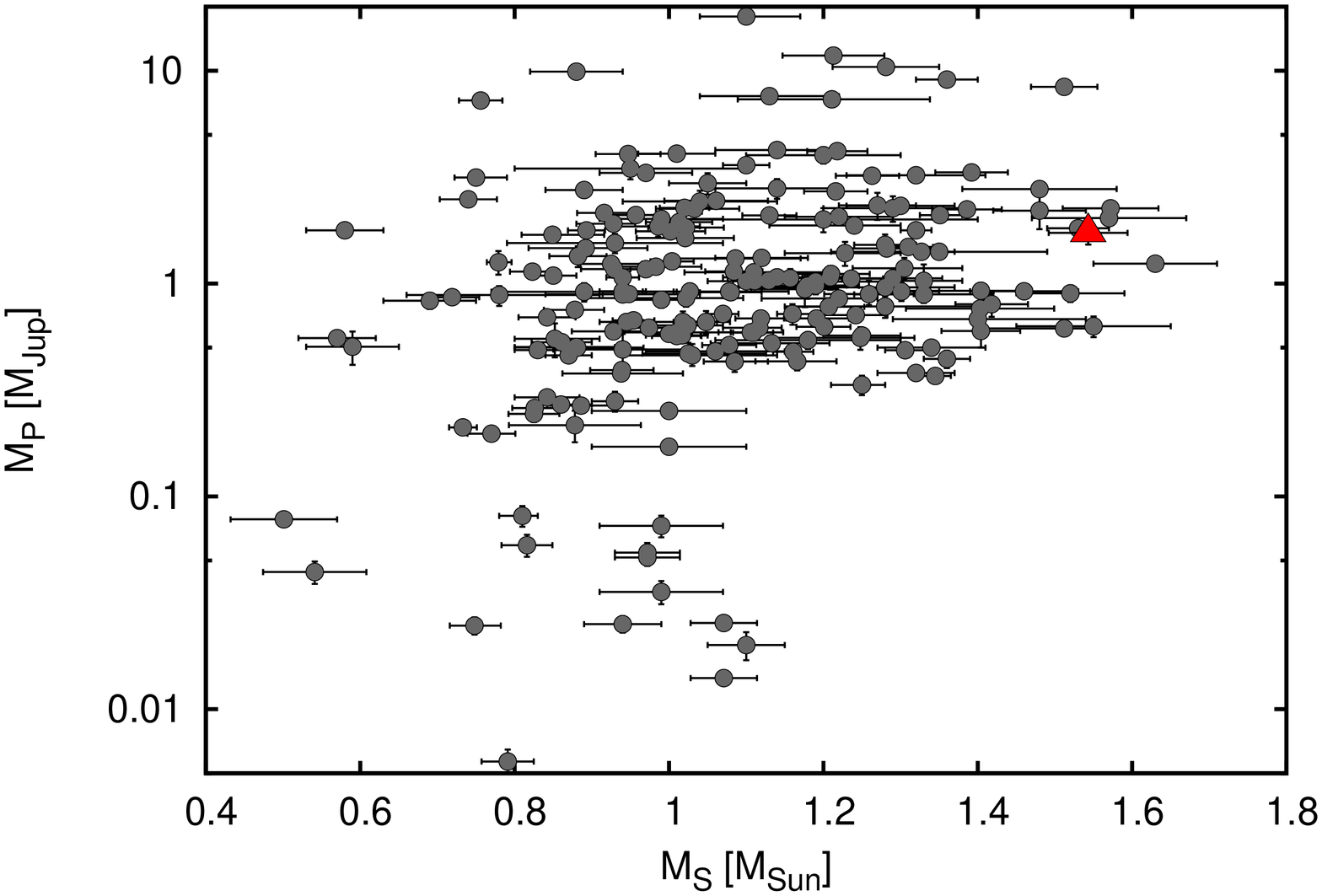}
\caption[]{
   Planet mass as a function of stellar mass. All planets with mass, radius and host mass uncertainties all 
   less than $20 \%$ are included in this plot. 
    \hatcurb{} is noted by the red triangle.  

\label{fig:starmplanetm}}
\end{figure}

\begin{figure}[]
\plotone{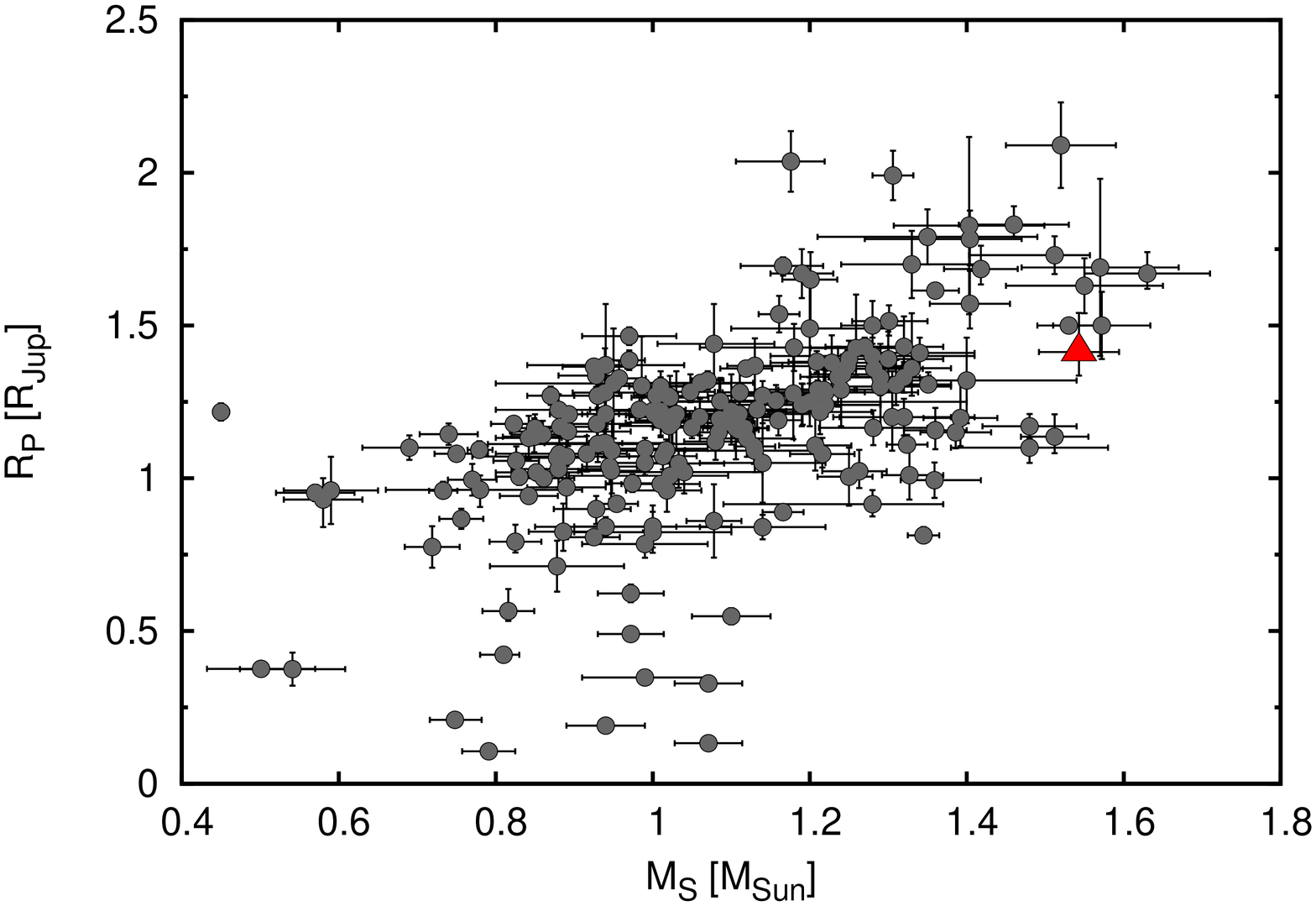}
\caption[]{
   Planet radius as a function of stellar mass. All planets with mass, radius and host mass uncertainties all 
   less than $20 \%$ are included in this plot. 
    \hatcurb{} is noted by the red triangle.  

\label{fig:starmplanetr}}
\end{figure}
\begin{figure}[]
\plotone{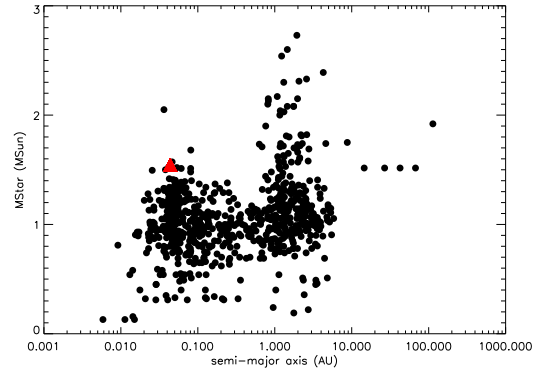}
\caption[]{
   Stellar mass as a function of semi-major Axis of all known exoplanets. \hatcurb{} is indicated with a red triangle.
\label{fig:au}}
\end{figure}
\ifthenelse{\boolean{emulateapj}}{
  \begin{deluxetable*}{lccr}
}{
  \begin{deluxetable}{lccr}
}
\tablewidth{0pc}
\tabletypesize{\scriptsize}
\tablecaption{
    Stellar Parameters for \hatcur{} 
    \label{tab:stellar}
}
\tablehead{
    \multicolumn{1}{c}{} &
    \multicolumn{1}{c}{{\bf Fixed Circ.}\tablenotemark{a}} &
    \multicolumn{1}{c}{Free Eccen.} &
    \multicolumn{1}{c}{} \\
    \multicolumn{1}{c}{~~~~~~~~Parameter~~~~~~~~} &
    \multicolumn{1}{c}{\bf Value}                     &
    \multicolumn{1}{c}{Value}                     &
    \multicolumn{1}{c}{Source}    
}
\startdata
\noalign{\vskip -3pt}
\sidehead{Identifying Information}
~~~~R.A. (h:m:s)                      &  \hatcurCCra{} & $\ldots$ & 2MASS\\
~~~~Dec. (d:m:s)                      &  \hatcurCCdec{} & $\ldots$ & 2MASS\\
~~~~HD ID                             &  \hatcurCChd{} & $\ldots$ & HD\\
~~~~GSC ID                            &  \hatcurCCgsc{} & $\ldots$ & GSC\\
~~~~2MASS ID                          &  \hatcurCCtwomass{} & $\ldots$ & 2MASS\\
\sidehead{Spectroscopic properties}
~~~~$\teffstar$ (K)\dotfill         &  \hatcurSMEteff{} & \hatcurSMEteffeccen{} & SPC\tablenotemark{b}\\
~~~~$\feh$\dotfill                  &  \hatcurSMEzfeh{} & \hatcurSMEzfeheccen{} & SPC                 \\
~~~~$\vsini$ (\kms)\dotfill         &  \hatcurSMEvsin{} & \hatcurSMEvsineccen{} & SPC                 \\
~~~~$\gamma_{\rm RV}$ (\kms)\dotfill&  \hatcurTRESgamma{} & $\ldots$ & TRES                  \\
\sidehead{Photometric properties}
~~~~$B$ (mag)\dotfill               &  \hatcurCCtassmB{} & $\ldots$ & APASS                \\
~~~~$V$ (mag)\dotfill               &  \hatcurCCtassmv{} & $\ldots$ & APASS               \\
~~~~$I$ (mag)\dotfill               &  \hatcurCCtassmI{} & $\ldots$ & TASS               \\
~~~~$g$ (mag)\dotfill               &  \hatcurCCtassmg{} & $\ldots$ & APASS                \\
~~~~$r$ (mag)\dotfill               &  \hatcurCCtassmr{} & $\ldots$ & APASS                \\
~~~~$i$ (mag)\dotfill               &  \hatcurCCtassmi{} & $\ldots$ & APASS                \\
~~~~$J$ (mag)\dotfill               &  \hatcurCCtwomassJmag{} & $\ldots$ & 2MASS           \\
~~~~$H$ (mag)\dotfill               &  \hatcurCCtwomassHmag{} & $\ldots$ & 2MASS           \\
~~~~$K_s$ (mag)\dotfill             &  \hatcurCCtwomassKmag{} & $\ldots$ & 2MASS           \\
\sidehead{Derived properties}
~~~~$\mstar$ ($\msun$)\dotfill      &  \hatcurISOmlong{} & \hatcurISOmlongeccen{} & \hatcurisoshort{}+\hatcurlumind{}+SPC\tablenotemark{c}\\
~~~~$\rstar$ ($\rsun$)\dotfill      &  \hatcurISOrlong{} & \hatcurISOrlongeccen{} & \hatcurisoshort{}+\hatcurlumind{}+SPC         \\
~~~~$\loggstar$ (cgs)\dotfill       &  \hatcurISOlogg{} & \hatcurISOloggeccen{} & \hatcurisoshort{}+\hatcurlumind{}+SPC         \\
~~~~$\lstar$ ($\lsun$)\dotfill      &  \hatcurISOlum{} & \hatcurISOlumeccen{} & \hatcurisoshort{}+\hatcurlumind{}+SPC         \\
~~~~$M_V$ (mag)\dotfill             &  \hatcurISOmv{} & \hatcurISOmveccen{} & \hatcurisoshort{}+\hatcurlumind{}+SPC         \\
~~~~$M_K$ (mag,\hatcurjhkfilset{})&  \hatcurISOMK{} & \hatcurISOMKeccen{} & \hatcurisoshort{}+\hatcurlumind{}+SPC         \\
~~~~Age (Gyr)\dotfill               &  \hatcurISOage{} & \hatcurISOageeccen{} & \hatcurisoshort{}+\hatcurlumind{}+SPC         \\
~~~~$A_{V}$ (mag)\tablenotemark{d}\dotfill           &  \hatcurXAv{} & \hatcurXAveccen{} & \hatcurisoshort{}+\hatcurlumind{}+SPC\\
~~~~Distance (pc)\dotfill           &  \hatcurXdistred{} & \hatcurXdistredeccen{} & \hatcurisoshort{}+\hatcurlumind{}+SPC\\
\enddata
\tablenotetext{a}{
    We adopt parameters from the model where the eccentricity is fixed
    to zero. The Bayesian evidence ratio strongly favors this model
    over a model where the eccentricity is allowed to vary. See
    \refsecl{analysis}.
}
\tablenotetext{b}{
    SPC = ``Stellar Parameter Classification'' method based on
    cross-correlating high-resolution spectra against synthetic
    templates \citep{buchhave:2012}. These parameters rely primarily
    on SPC, but have a small dependence also on the iterative analysis
    incorporating the isochrone search and global modeling of the
    data, as described in the text.  } 
\tablenotetext{c}{
    \hatcurisoshort{}+\hatcurlumind{}+SPC = Based on the
    \hatcurisoshort{}\ isochrones \citep{yi:2001},
    \hatcurlumind{}\ (directly related to stellar density) as a
    luminosity indicator, and the SPC results.
} 
\tablenotetext{d}{ Total \band{V} extinction to the star determined
  by comparing the catalog broad-band photometry listed in the table
  to the expected magnitudes from the
  \hatcurisoshort{}+\hatcurlumind{}+SPC model for the star. We use the
  \citet{cardelli:1989} extinction law.  }
\ifthenelse{\boolean{emulateapj}}{
  \end{deluxetable*}
}{
  \end{deluxetable}
}
\ifthenelse{\boolean{emulateapj}}{
  \begin{deluxetable*}{lcc}
}{
  \begin{deluxetable}{lcc}
}
\tabletypesize{\scriptsize}
\tablecaption{Parameters for the transiting planet \hatcurb{}. We list the adopted parameters that result from assuming a fixed circular orbit, together with those that result from allowing the eccentricity to vary.\label{tab:planetparam}}
\tablehead{
    \multicolumn{1}{c}{} &
    \multicolumn{1}{c}{{\bf Fixed Circ.}\tablenotemark{a}} &
    \multicolumn{1}{c}{Free Eccen.} \\
    \multicolumn{1}{c}{~~~~~~~~Parameter~~~~~~~~} &
    \multicolumn{1}{c}{\bf Value}                     &
    \multicolumn{1}{c}{Value}                     
}
\startdata
\noalign{\vskip -3pt}
\sidehead{\Lc{} parameters}
~~~$P$ (days)             \dotfill    & $\hatcurLCP{}$              & $\hatcurLCPeccen{}$             \\
~~~$T_c$ (${\rm BJD}$)    
      \tablenotemark{b}   \dotfill    & $\hatcurLCT{}$              & $\hatcurLCTeccen{}$              \\
~~~$T_{14}$ (days)
      \tablenotemark{b}   \dotfill    & $\hatcurLCdur{}$            & $\hatcurLCdureccen{}$            \\
~~~$T_{12} = T_{34}$ (days)
      \tablenotemark{b}   \dotfill    & $\hatcurLCingdur{}$         & $\hatcurLCingdureccen{}$         \\
~~~$\arstar$              \dotfill    & $\hatcurPPar{}$             & $\hatcurPPareccen{}$             \\
~~~$\zrstar$\tablenotemark{c}              \dotfill    & $\hatcurLCzeta{}$\phn       & $\hatcurLCzetaeccen{}$\phn       \\
~~~$\rpl/\rstar$          \dotfill    & $\hatcurLCrprstar{}$        & $\hatcurLCrprstareccen{}$        \\
~~~$b^2$                  \dotfill    & $\hatcurLCbsq{}$            & $\hatcurLCbsqeccen{}$            \\
~~~$b \equiv a \cos i/\rstar$
                          \dotfill    & $\hatcurLCimp{}$           & $\hatcurLCimpeccen{}$            \\
~~~$i$ (deg)              \dotfill    & $\hatcurPPi{}$\phn         & $\hatcurPPieccen{}$\phn          \\

\sidehead{Limb-darkening coefficients \tablenotemark{d}}
~~~$c_1,z$ (linear term)  \dotfill    & $\hatcurLBiz{}$            & $\hatcurLBizeccen{}$             \\
~~~$c_2,z$ (quadratic term) \dotfill  & $\hatcurLBiiz{}$           & $\hatcurLBiizeccen{}$            \\
~~~$c_1,r$               \dotfill    & $\hatcurLBir{}$             & $\hatcurLBireccen{}$             \\
~~~$c_2,r$               \dotfill    & $\hatcurLBiir{}$            & $\hatcurLBiireccen{}$            \\

\sidehead{RV parameters}
~~~$K$ (\ms)              \dotfill    & $\hatcurRVK{}$\phn\phn      & $\hatcurRVKeccen{}$\phn\phn      \\
~~~$\sqrt{e} \cos \omega$ 
                          \dotfill    & $0$ (fixed) & $\hatcurRVrkeccen{}$\phs          \\
~~~$\sqrt{e} \sin \omega$
                          \dotfill    & $0$ (fixed) & $\hatcurRVrheccen{}$              \\
~~~$e \cos \omega$ 
                          \dotfill    & $0$ (fixed) & $\hatcurRVkeccen{}$\phs          \\
~~~$e \sin \omega$
                          \dotfill    & $0$ (fixed) & $\hatcurRVheccen{}$              \\
~~~$e$                    \dotfill    & $0$ (fixed) & $\hatcurRVecceneccen{}$          \\
~~~$\omega$ (deg)         \dotfill    & $\ldots$ & $\hatcurRVomegaeccen{}$\phn      \\
~~~RV jitter FLWO~1.5\,m/TRES (\ms)\tablenotemark{e}        \dotfill    & \hatcurRVjitterA{}           & \hatcurRVjitterAeccen{}           \\

~~~RV jitter OHP~1.93\,m/SOPHIE (\ms)\tablenotemark{e}        \dotfill    & \hatcurRVjitterB{}         & \hatcurRVjitterBeccen{}           \\

\sidehead{Planetary parameters}
~~~$\mpl$ ($\mjup$)       \dotfill    & $\hatcurPPmlong{}$          & $\hatcurPPmlongeccen{}$          \\
~~~$\rpl$ ($\rjup$)       \dotfill    & $\hatcurPPrlong{}$          & $\hatcurPPrlongeccen{}$          \\
~~~$C(\mpl,\rpl)$
    \tablenotemark{f}     \dotfill    & $\hatcurPPmrcorr{}$         & $\hatcurPPmrcorreccen{}$         \\
~~~$\rhopl$ (\gcmc)       \dotfill    & $\hatcurPPrho{}$            & $\hatcurPPrhoeccen{}$            \\
~~~$\log g_p$ (cgs)       \dotfill    & $\hatcurPPlogg{}$           & $\hatcurPPloggeccen{}$           \\
~~~$a$ (AU)               \dotfill    & $\hatcurPParel{}$          & $\hatcurPPareleccen{}$           \\
~~~$T_{\rm eq}$ (K)\tablenotemark{g}        \dotfill   & $\hatcurPPteff{}$           & $\hatcurPPteffeccen{}$           \\
~~~$\Theta$\tablenotemark{h} \dotfill & $\hatcurPPtheta{}$         & $\hatcurPPthetaeccen{}$          \\
~~~$\langle F \rangle$ ($10^{9}$\ergscmsq) \tablenotemark{i}
                          \dotfill    & $\hatcurPPfluxavg{}$       & $\hatcurPPfluxavgeccen{}$        \\ [-1.5ex]
\enddata
\tablenotetext{a}{
    Adopted parameters are taken from a model where the eccentricity
    is fixed to zero as this model is strongly preferred by the
    Bayesian evidence ratio over a model where the eccentricity is
    allowed to vary. Parameters that result from the model where the
    eccentricity is allowed to vary are displayed in the subsequent
    column for reference. See \refsecl{analysis}.
}
\tablenotetext{b}{
    Reported times are in Barycentric Julian Date calculated directly
    from UTC, {\em without} correction for leap seconds.
    \ensuremath{T_c}: Reference epoch of mid transit that
    minimizes the correlation with the orbital period.
    \ensuremath{T_{14}}: total transit duration, time
    between first to last contact;
    \ensuremath{T_{12}=T_{34}}: ingress/egress time, time between first
    and second, or third and fourth contact.
}
\tablenotetext{c}{
    Reciprocal of the half duration of the transit used as a jump
    parameter in our MCMC analysis in place of $\arstar$. It is
    related to $\arstar$ by the expression $\zrstar = \arstar
    (2\pi(1+e\sin \omega))/(P \sqrt{1 - b^{2}}\sqrt{1-e^{2}})$
    \citep{bakos:2010:hat11}.
}
\tablenotetext{d}{
    Values for a quadratic law, adopted from the tabulations by
    \cite{claret:2004} according to the spectroscopic (SPC) parameters
    listed in \reftabl{stellar}.
}
\tablenotetext{e}{
    Error term, either astrophysical or instrumental in origin, added
    in quadrature to the formal RV errors for the listed
    instrument. This term is varied in the fit assuming a prior inversely 
    proportional to the jitter.
}
\tablenotetext{f}{
    Correlation coefficient between the planetary mass \mpl\ and
    radius \rpl\ determined from the parameter posterior distribution
    via $C(\mpl,\rpl) = <(\mpl - <\mpl>)(\rpl -
    <\rpl>)>/(\sigma_{\mpl}\sigma_{\rpl})>$ where $< \cdot >$ is the
    expectation value operator, and $\sigma_x$ is the standard
    deviation of parameter $x$.
}
\tablenotetext{g}{
    Planet equilibrium temperature averaged over the orbit, calculated
    assuming a Bond albedo of zero, and that flux is reradiated from
    the full planet surface.
}
\tablenotetext{h}{
    The Safronov number is given by $\Theta = \frac{1}{2}(V_{\rm
    esc}/V_{\rm orb})^2 = (a/\rpl)(\mpl / \mstar )$
    \citep[see][]{hansen:2007}.
}
\tablenotetext{i}{
    Incoming flux per unit surface area, averaged over the orbit.
}
\ifthenelse{\boolean{emulateapj}}{
  \end{deluxetable*}
}{
  \end{deluxetable}
}
%



\section{Discussion}
\label{sec:discussion}

We have presented the discovery of \hatcurb{}, a hot Jupiter orbiting  
one of the most massive stars to have \mpl\ and \rpl\ accurately determined.
Figure~\ref{fig:starmplanetm} and Figure~\ref{fig:starmplanetr} show planet mass
as a function of stellar mass and planet radius as a function of stellar mass, respectively, of all known TEPs with 
mass and radius determined to $20 \%$. Not only is this one of the most massive planets to 
orbit an intermediate-mass (IM) star (\mstar\ $\ge$ $1.5$ \msun), but \hatcurb{} orbits very close to its host star at a
semi-major axis of \hatcurPParel{} AU. There are very few known massive planets with orbital distances $< 0.1$ AU
around $242$ IM stars. Doppler studies \citep[][]{johnson:2007,lovismayor:2007} have targeted
IM evolved stars specifically because main sequence IM stars have high stellar jitter and few Doppler absorption lines
due to rotational broadening \citep[][]{galland:2005}, whereas IM subgiants and giants have lower jitter and more Doppler lines
due to slower rotation and a cooler photosphere.  
We are finding that IM stars that host hot Jupiters $< 0.1$ AU (e.g., HAT-P-40b, WASP-79b, Kepler-14b, HAT-P-7b, etc.) are mainly being discovered by transit searches
\citep[][]{hartman:2012:hat39hat41,smalley:2012:wasp78wasp79,buchhave:2011:kepler14,pal:2008:hat7}. Up until recently, planets with 
semi-major axes $< 0.6$ AU orbiting stars with masses $> 1.5$ \msun\ were non-existent and this area was termed
the ``planet desert'' by \cite{bowler:2010}. Still, there are only a handful of planets around IM stars with a semi-major axis $< 0.1$ AU and
while there are many more that have been discovered orbiting at $> 0.6$ AU, there is still a gap between $ 0.1$ AU and $0.6$ AU where
we have not discovered any planets in the IM regime (see Figure~\ref{fig:au}). This does not seem to be an observational bias as massive planets would have detectable RV signals.
Recently, however, the masses of the subgiant stars targeted by these surveys have been called into question \citep{lloyd:2011,lloyd:2013,schlaufman:2013} with suggestions that they have been systematically overestimated. Even the planetary nature of the periodic RV variations has been called into question \citep{lloyd:2013}. The matter is a subject of current debate \citep{johnson:2013}.
Continued searches for planets around IM stars are essential to garner a true understanding of the nature of planet formation and evolution
of hot Jupiter planets.

Two of the TRES spectra were obtained during transit and were not included in the model because of the risk
that the Doppler velocities derived from those observations were distorted by possible Rossiter-McLaughlin effects.
This Rossiter-McLaughlin (R-M) effect can be used to determine the sky-projected angle
between a planet's orbital axis and the star's rotation axis.
It has been shown that hot Jupiters have a wide
range of these projected angles \citep[][]{albrecht:2012}, telling us that a planetary orbit
can be misaligned with the stellar rotation and can even be retrograde. An R-M analysis can 
also provide an independent measure of \vsini.
\hatcur{} is a strong candidate to consider for R-M follow-up due to the brightness of the host star, 
as well as it's rapid rotation of $16$ {\kms}. This latter factor increases the amplitude of the signal,
however it also tends to increase the RV measurement uncertainty. 
We note that the two spectra taken during transit show deviations in the opposite direction of
that expected for alignment, suggesting that this system may be misaligned.  However, because of
the large velocity jitter that we needed to model this system and the fact that the R-M amplitude
is about twice what we would expect, the two observations taken during transit 
can not be trusted, and continuous velocity monitoring during one or more transits is needed for
a proper R-M experiment. 
With the expected maximum R-M amplitude of $94$ 
{\ms} and the brightness of the host star, this would be a good target for future R-M follow-up, 
even with a modest sized telescope, and an important addition
to the collection of planets with R-M studies. 
In addition, the independent determination
of \vsini\ from a R-M follow-up would be an excellent additional constraint in determining
better stellar parameter classification. 


\acknowledgements 

\paragraph{Acknowledgements}
HATNet operations have been funded by NASA grants NNG04GN74G and
NNX13AJ15G. Follow-up of HATNet targets has been
partially supported through NSF grant AST-1108686. G.\'A.B., Z.C. and
K.P. acknowledge partial support from NASA grant NNX09AB29G. 
K.P. acknowledges support from NASA grant NNX13AQ62G.
GT acknowledges partial support from NASA grant NNX09AF59G. We
acknowledge partial support also from the Kepler Mission under NASA
Cooperative Agreement NCC2-1390 (D.W.L., PI). G.K.~thanks the
Hungarian Scientific Research Foundation (OTKA) for support through
grant K-81373. 
The authors thank all the staff of Haute-Provence Observatory for their 
contribution to the success of the SOPHIE project and their support at the 
1.93-m telescope. IB acknowledges the support of the European Research Council
European Community under the FP7 through a Starting Grant, as well from Fundaccao 
para a Ciência e a Tecnologia (FCT), Portugal, through SFRH/BPD/81084/2011 an 
the project PTDC/CTE- AST/098528/2008.
Data presented in this paper are based on observations
obtained at the HAT station at the Submillimeter Array of SAO, and the
HAT station at the Fred Lawrence Whipple Observatory of SAO. The
authors wish to recognize and acknowledge the very significant
cultural role and reverence that the summit of Mauna Kea has always
had within the indigenous Hawaiian community. We are most fortunate to
have the opportunity to conduct observations from this
mountain. 

\clearpage
\bibliographystyle{apj}
\bibliography{htrbib.bib}

\end{document}